\documentclass[12pt]{article}
\pdfoutput=1
\usepackage{amsmath}
\usepackage{graphics,graphicx,xcolor}

\def\be{\begin{equation}}
\def\ee{\end{equation}}
\def\arr{\begin{array}{rll}}
\def\ea{\end{array}}
\def\bea{\begin{eqnarray}}
\def\eea{\end{eqnarray}}

\def\N2{$N{=}2$}

\def\>{\rangle}
\def\<{\langle}
\def\+{\dagger}
\def\={\ =\ }

\def\bal{\begin{aligned}}
\def\eal{\end{aligned}}

\begin{document}
\begin{titlepage}
\setcounter{page}{0}
\begin{center}
{\LARGE\bf  Equations of fluid mechanics with}\\
\vskip 0.5cm
{\LARGE\bf    $\mathcal{N}=1$ Schr\"odinger supersymmetry  }\\
\vskip 1.5cm
\textrm{\Large Anton Galajinsky \ }
\vskip 0.7cm
{\it
Laboratory of Applied Mathematics and Theoretical Physics, \\ TUSUR, Lenin ave. 40, 634050 Tomsk, Russia} \\

\vskip 0.2cm
{e-mail: a.galajinsky@tusur.ru}
\vskip 0.5cm
\end{center}

\begin{abstract} \noindent
Equations of fluid mechanics with 
$\mathcal{N}=1$ Schr\"odinger supersymmetry are formulated
within the method of nonlinear realizations of Lie groups.
\end{abstract}

\vspace{0.5cm}

\end{titlepage}
\renewcommand{\thefootnote}{\arabic{footnote}}
\setcounter{footnote}0

\noindent
{\bf 1. Introduction}\\

Supersymmetric extensions of nonrelativistic fluid mechanics are of interest 
for several reasons. One the one hand, systems of such a type in $(1+1)$ and $(1+2)$
dimensions are tightly connected with superstring and supermembrane 
theories \cite{JP,BJ}. On the other hand, some of them admit
an infinite set of conserved charges in
involution, thus furnishing nontrivial examples of integrable field 
theories \cite{DP}. They also provide a convenient framework for 
developing solution--generating techniques \cite{FKT,GH}.

A group--theoretic description of fluid mechanics in terms of group--valued variables was 
proposed in \cite{BJLNP}. It was designed so as to take into account constituent 
particles which carry nonabelian charges as well as for incorporating anomalies 
(for a review and further references see \cite{JNPP}). 

An alternative approach, which relied upon the method of nonlinear 
realizations of Lie groups \cite{CWZ}, was studied in a recent work \cite{AG}.
The continuity equation and
the Euler equation were constructed directly in terms of the
Maurer--Cartan invariants, which ensured a desired symmetry. 
Within this method, perfect fluid equations with the Schr\"odinger symmetry, or
the $\ell$--conformal Galilei symmetry, or the Lifshitz symmetry, 
as well as a proper equation of state, come about quite naturally \cite{AG} 
(for related earlier developments see \cite{HH}--\cite{HZ}).

In connection with the previous studies of supersymmetric fluid mechanics, 
it is natural to wonder whether the procedure in \cite{AG} can be generalized so as to 
incorporate supersymmetry transformations. The goal of this work is to investigate 
the case of the $\mathcal{N}=1$ Schr\"odinger supergroup \cite{GGT}. 
Note that dynamical realizations of (super) Schr\"odinger group are of potential 
interest from the standpoint of the nonrelativistic holography \cite{BG}.  
Fluid mechanics with conformal symmetry is in the focus of the fluid/gravity 
correspondence \cite{MR}.

As far as realizations in superspace are concerned, in this work we choose 
the simplest option in which a single Grassmann--odd real variable $\theta$ 
is introduced as a superpartner of the temporal variable $t$. Together they form 
$\mathcal{R}^{(1|1)}$ superspace.
Cartesian coordinates $x_i$, $i=1,\dots,n$, are inert under supertranslations.
As a result, each physical field in question, like the density of a fluid or the velocity 
vector field, is accompanied by a superpartner of 
the opposite Grassmann parity. An essential new ingredient of the construction is an extra vector superfield, 
the presence of which is dictated by the superalgebra at hand.

The work is organized as follows. In the next section, 
structure relations of the $\mathcal{N}=1$ Schr\"odinger superalgebra are briefly reminded.
In Sect. 3 and Sect. 4, the method of nonlinear realizations is applied  
to build superfield equations of motion which govern fluid mechanics with the $\mathcal{N}=1$ Schr\"odinger supersymmetry.
The equations are given in terms of invariant Maurer--Cartan one--forms and two invariant derivatives, 
one of which is Grassmann--even and the other is Grassmann--odd. In the concluding Sect. 5, we summarize 
our results and discuss possible further developments.

Throughout the paper, summation over repeated indices is understood unless otherwise stated. 
A derivative with respect to a Grassmann--odd variable is assumed to be the left derivative.

\vspace{0.5cm}

\noindent
{\bf 2. $\mathcal{N}=1$ Schr\"odinger superalgebra}\\

Grassmann--even generators entering the Lie superalgebra associated with the $\mathcal{N}=1$ Schr\"odinger 
supergroup describe temporal translation ($H$), dilatation ($D$), and special
conformal transformation ($C$), which form $sl(2,R)$ subalgebra, as well as spatial translations 
($P_i$) and the Galilei boosts ($K_i$), where $i=1,\dots,n$ and $n$ is the spatial dimension.~\footnote{The algebra also 
involves spatial rotations, which in what follows will be disregarded.} 
They obey the structure relations 
\begin{align}\label{SA}
&
[H,D]={\rm i} H, &&  [H,C]=2 {\rm i} D, && [D,C]={\rm i} C, && [H,K_i]={\rm i} P_i,
\nonumber\\[2pt]
&
[D,P_i]=-\frac{{\rm i}}{2} P_i, && [D,K_i]=\frac{{\rm i}}{2} K_i, && [C,P_i]=-{\rm i} K_i. &&
\end{align}

Grassmann--odd generators include the $\mathcal{N}=1$ supersymmetry charge ($Q$), the generator of 
superconformal transformations ($S$), and the superpartner ($\Lambda_i$, $i=1,\dots,n$) of the Galilie boosts. 
These generators add to (\ref{SA}) the following structure relations 
\begin{align}\label{SSA}
&
\{Q,Q\}=2 H, &&  \{S,S\}=2 C, && \{Q,S \}=-2 D, && \{Q,\Lambda_i\}=P_i,
\nonumber\\[4pt]
&
\{S,\Lambda_i\}=-K_i, && [H,S]=-{\rm i} Q, && [C,Q]={\rm i} S, && [Q,K_i]={\rm i} \Lambda_i,
\nonumber\\[2pt]
&
[D,Q]=-\frac{{\rm i}}{2} Q, && [D,S]=\frac{{\rm i}}{2} S, && [S,P_i]={\rm i} \Lambda_i.
\end{align}
A remarkable property of the superalgebra is that brackets among the fermionic operators $(Q,S,\Lambda_i)$
produce all the bosonic generators $(H,D,C,P_i,K_i)$. As a consequence, when verifying 
the $\mathcal{N}=1$ Schr\"odinger supersymmetry of a specific physical model, 
it suffices to focus on the transformations associated with the Grassmann--odd generators only.

As far as realizations in superspace are concerned, in this work we choose
the simplest option
in which a single Grassmann--odd real variable $\theta$ is introduced. It is regarded as a superpartner 
of a temporal variable $t$. Together they form 
$\mathcal{R}^{(1|1)}$ superspace.
Cartesian coordinates $x_i$, $i=1,\dots,n$, are inert under supertranslations.
A realization of the superalgebra (\ref{SA}), (\ref{SSA}) in such a superspace reads
\begin{align}\label{gener}
&
H={\rm i} \frac{\partial}{\partial t}, &&
D={\rm i}  \left(t \frac{\partial}{\partial t}+\frac 12 x_i \frac{\partial}{\partial x_i} 
+\frac 12 \theta \frac{\partial}{\partial \theta}\right), &&
C={\rm i}  \left(t^2 \frac{\partial}{\partial t}+ 
t x_i \frac{\partial}{\partial x_i}+t \theta \frac{\partial}{\partial \theta}\right), 
\nonumber\\[2pt]
&
P_i={\rm i} \frac{\partial}{\partial x_i}, &&
K_i={\rm i} t \frac{\partial}{\partial x_i}, &&
Q=\frac{\partial}{\partial \theta}+{\rm i} \theta \frac{\partial}{\partial t}, 
\nonumber\\[2pt]
&
S=-t Q-{\rm i} \theta x_i \frac{\partial}{\partial x_i}, &&
\Lambda_i={\rm i} \theta \frac{\partial}{\partial x_i}. &&
\end{align}
The first appearance of the $\mathcal{N}=1$ Schr\"odinger 
superalgebra in mechanics was the model of
the non--relativistic spin--$\frac 12$ particle \cite{GGT} (see also the discussion in \cite{DH}).

\newpage

\noindent
{\bf 3. Coordinates, fields, and symmetry transformation rules}\\

The method of nonlinear realizations \cite{CWZ} provides a convenient 
framework for building dynamical systems with a given symmetry group, which relies 
solely upon structure relations of the corresponding Lie algebra.
It includes several steps. 
Firstly, one introduces coordinates and fields, 
whose number is in general equal to the number of generators in 
a Lie algebra at hand.
Then one builds a group--theoretic element $g$, which is a product 
of the exponentials $e^{{\rm i} a T}$, 
where $a$ is a coordinate (or a field) and $T$ is a Lie algebra generator. 
Then one examines the action of the group upon 
the group--theoretic element $g$, which is realized by the left multiplication, 
and finds out infinitesimal transformation 
rules for the coordinates and fields. As the final step, one builds 
the Maurer-Cartan one--forms $g^{-1} d g$, which are automatically invariant
under the left action of the group upon the group--theoretic element. They furnish the building blocks for constructing
invariant equations of motion (for more details including specific examples see Sect. II in \cite{AG}). 

Turning to the $\mathcal{N}=1$ Schr\"odinger superalgebra, it seems natural to link the coordinates $t$, $\theta$, and 
$x_i$ to the corresponding (super)translation generators $H$, $Q$, and $P_i$, 
respectively.
The remaining generators 
entering the superalgebra are 
accompanied by as yet unspecified superfields of the same Grassmann parity. A group--theoretic element which is 
central for our consideration reads
\be\label{gte}
g=e^{{\rm i} t H} e^{\theta Q} e^{\chi(t,\theta,x) S} e^{\psi_i (t,\theta,x) \Lambda_i} 
e^{{\rm i} x_i P_i} 
e^{{\rm i}  \upsilon_i (t,\theta,x) K_i} 
e^{{\rm i} u(t,\theta,x) D} e^{{\rm i} w(t,\theta,x) C}.
\ee
In particular, $\chi$, $\psi_i$ are Grassmann--odd real superfields, 
while $\upsilon_i$, $u$, and $w$ are Grassmann--even real superfields.
Note that the factors contributing to the group--theoretic element in general do not commute. 
The choice (\ref{gte}) proves to be most convenient for practical calculations.

As the next step, one has to determine how the coordinates and superfields transform under 
the left action of the $\mathcal{N}=1$ Schr\"odinger supergroup upon the group--theoretic element 
(infinitesimal parameters associated with the generators are designated by 
similar lowercase letters)
\bea\label{CT}
&&
g_0 \cdot g=
e^{{\rm i} h H} e^{q Q} e^{s S} e^{\lambda_i  \Lambda_i} e^{{\rm i} p_i P_i} 
e^{{\rm i}  k_i  K_i} 
e^{{\rm i} d D} e^{{\rm i} c C}
\cdot g=
\nonumber\\[2pt]
&&
e^{{\rm i} t' H} e^{\theta' Q} e^{\chi'(t',\theta',x') S} 
e^{\psi'_i (t',\theta',x') \Lambda_i} e^{{\rm i} x'_i P_i} 
e^{{\rm i}  \upsilon'_i (t',\theta',x') K_i} 
e^{{\rm i} u'(t',\theta',x') D} e^{{\rm i} w'(t',\theta',x') C}.
\eea
Making use of the Baker--Campbell--Hausdorff formulae 
\bea\label{BCH}
&&
e^{A} e^{B}=e^{A+B+\frac 12 [A,B]+\frac{1}{12} [A,[A,B]+\frac{1}{12} [B,[B,A]+\dots},
\nonumber\\[2pt]
&&
e^{iA}~ T~ e^{-iA}=T+\sum_{n=1}^\infty\frac{i^n}{n!}
\underbrace{[A,[A, \dots [A,T] \dots]]}_{n~\rm times},
\eea
one gets (each single transformation is separated by a semicolon)
\begin{align}\label{TRL}
&
t'=t+h; &&
\nonumber\\[6pt]
&
t'=t+{\rm i} q \theta, \quad \theta'=\theta+q; &&
\nonumber\\[6pt]
&
t'=t\left(1-{\rm i} s \theta \right), && 
\chi'(t',\theta',x')=\chi(t,\theta,x)+s \left(1+{\rm i} \theta\chi (t,\theta,x) \right),
\nonumber\\[2pt]
& \theta'=\theta-t s,
&& \upsilon'_i (t',\theta',x')=\upsilon_i (t,\theta,x) \left(1+{\rm i} s \theta\right)
+{\rm i} x_i  s \chi (t,\theta,x),
\nonumber\\[2pt]
& x'_i=x_i \left(1-{\rm i} s \theta \right),
&& u'(t',\theta',x')=u (t,\theta,x)-2 {\rm i} s \theta,
\nonumber\\[2pt]
&
&& w'(t',\theta',x')=w (t,\theta,x)+e^{u (t,\theta,x)} {\rm i} s \chi (t,\theta,x);
\nonumber\\[6pt]
&
x'_i=x_i+{\rm i} \lambda_i \theta, &&
\psi'_i (t',\theta',x')=\psi_i (t,\theta,x)+\lambda_i 
\left(1+{\rm i} \theta \chi (t,\theta,x)\right),
\nonumber\\[2pt]
&
&& \upsilon'_i (t',\theta',x')=\upsilon_i (t,\theta,x) -{\rm i} \lambda_i \chi (t,\theta,x);
\nonumber\\[6pt]
&
x'_i=x_i+p_i, && \psi'_i (t',\theta',x')=\psi_i (t,\theta,x)+p_i \chi(t,\theta,x);
\nonumber\\[6pt]
&
x'_i=x_i+t k_i, && \psi'_i (t',\theta',x')=\psi_i (t,\theta,x)
+k_i \left(\theta+t \chi(t,\theta,x) \right),
\nonumber\\[2pt]
&
&&
\upsilon'_i (t',\theta',x')=\upsilon_i (t,\theta,x)
+k_i \left(1-{\rm i} \theta \chi (t,\theta,x)\right);
\nonumber\\[6pt]
&
t'=t \left(1+d\right), && \chi' (t',\theta',x')=\chi(t,\theta,x) 
\left(1-d/2 \right),
\nonumber\\[2pt]
&
\theta'=\theta \left(1+d/2 \right), 
&& \upsilon'_i (t',\theta',x')=\upsilon_i (t,\theta,x) 
\left(1-d/2 \right),
\nonumber\\[2pt]
&
x'_i=x_i \left(1+d/2 \right), 
&& u' (t',\theta',x')=u(t,\theta,x)+d;
\nonumber\\[6pt]
&
t'=t \left(1+c t\right), && \chi' (t',\theta',x')=\chi (t,\theta,x)\left(1- ct\right) - c\theta,
\nonumber\\[2pt]
&
\theta'=\theta \left(1+ct \right),
&& \upsilon'_i (t',\theta',x')=
\upsilon_i (t,\theta,x) \left(1- ct \right)
+x_i c \left(1-{\rm i} \theta \chi (t,\theta,x) \right),
\nonumber\\[2pt]
&
x'_i=x_i \left(1+c t \right),
&& u' (t',\theta',x')=u(t,\theta,x)+2 c t,
\nonumber\\[2pt]
&
&& w' (t',\theta',x')=w(t,\theta,x)+e^{u(t,\theta,x)} c 
\left( 1-{\rm i} \theta \chi (t,\theta,x) \right).
\end{align}
In the relations above, we omitted coordinates which stay intact and superfields which transform as scalars. 
For example, in the unfolded form the first line in (\ref{TRL}) would read
\begin{align}
&
t'=t+h, && \theta'=\theta, && x'_i=x_i, 
\nonumber\\[2pt]
&
\chi'(t',\theta',x')=\chi(t,\theta,x), &&
\psi'_i (t',\theta',x')=\psi_i (t,\theta,x), 
&&
\upsilon'_i (t',\theta',x')=\upsilon_i (t,\theta,x),
\nonumber\\[2pt]
&
u'(t',\theta',x')=u(t,\theta,x), && w'(t',\theta',x')=w(t,\theta,x). &&
\end{align}

Representing the coordinate and field transformations above in the form $A'=A+\delta A$, where
\begin{align}
&
\delta_h=-{\rm i} h H, && \delta_q=q Q, && \delta_s=s S, &&
\delta_\lambda=\lambda_i \Lambda_i, 
\nonumber\\[2pt]
&
\delta_p=-{\rm i} p_i P_i, 
&& \delta_k=-{\rm i} k_i K_i, && \delta_d=-{\rm i} d D, && \delta_c=-{\rm i} c C,
\end{align}
and computing the commutators $[\delta_1,\delta_2]$, one can verify that 
the structure relations of the $\mathcal{N}=1$ Schr\"odinger superalgebra (\ref{SA}), (\ref{SSA}) are 
precisely reproduced.

\vspace{0.5cm}

\noindent
{\bf 4. Invariant one--forms, invariant derivatives, and invariant field equations}\\

As was mentioned above, given a group--theoretic element $g$, the Maurer--Cartan one--forms 
$g^{-1} d g$ hold invariant under 
the transformation $g'=g_0 \cdot g$, where $g_0$ is a group element chosen in an 
appropriate neighbourhood of the unit element. Specifying to $g$ in (\ref{gte}) and $g_0$ in (\ref{CT}), 
repeatedly using the Baker--Campbell--Hausdorff formulae (\ref{BCH}), and 
introducing the notation
\be
g^{-1} d g={\rm i} \omega_H  H +{\rm i} \omega_D D +{\rm i} \omega_C C +{\rm i} \omega_{P i} P_i  
+{\rm i} \omega_{K i} K_i+\omega_Q Q+\omega_S S+\omega_{\Lambda i} \Lambda_i,
\ee
one obtains 
\bea\label{MC}
&&
\omega_H=e^{-u} d \tau, \qquad \omega_Q=e^{-\frac{u}{2}} \left(d \theta+ \chi d \tau \right),
\qquad
\omega_D=du-2 w \omega_H+2 {\rm i} e^{\frac{u}{2}} \chi \omega_Q, 
\nonumber\\[2pt]
&&
\omega_S=e^{\frac{u}{2}} d \chi+w \omega_Q, \qquad 
\omega_C=dw-w du+ w^2 \omega_H-{\rm i} e^{\frac{u}{2}} \chi \left(\omega_S+w \omega_Q \right),
\nonumber\\[2pt]
&&
\omega_{\Lambda i}=d \psi_i-x_i d \chi-e^{\frac{u}{2}} \upsilon_i \omega_Q, \qquad
\omega_{P i}=e^{-\frac{u}{2}}\left(dx_i- \upsilon_i d \tau \right)-
{\rm i} \left(\psi_i-x_i \chi \right) \omega_Q,
\nonumber\\[2pt]
&&
\omega_{K i}=e^{\frac{u}{2}} \left(d \upsilon_i +{\rm i} \left(\psi_i-x_i \chi  \right) d\chi \right)-w \omega_{P i}
-{\rm i} e^{u}\upsilon_i \chi \omega_Q,
\eea
where we denoted
\be
d \tau=dt -{\rm i } \theta d \theta.
\ee
Note that under the infinitesimal dilatations, 
special conformal transformations, and the superconformal boosts $d \tau$ scales as 
\be
d \tau'=(1+d) d \tau, \qquad
\qquad d \tau'=(1+2ct) d \tau, \qquad d \tau'=(1-2 {\rm i} s \theta) d \tau,
\ee
while it remains intact
for other transformations from the 
$\mathcal{N}=1$ Schr\"odinger supergroup.

Taking into account the identity\footnote{Recall that throughout the text derivatives with respect to 
Grassmann--odd variables are the left derivatives.}
\be
\frac{\partial z'^A}{\partial z^B} \frac{\partial}{\partial z'^A}=\frac{\partial}{\partial z^B},
\ee
where $z^A=(t,\theta,x_i)$, and the transformation laws (\ref{TRL}), one can also verify 
that the following differential operators
\be\label{ID}
\nabla_i=e^{\frac{u}{2}} \frac{\partial}{\partial x_i}, \qquad 
\nabla=e^{\frac{u}{2}}  \left(\frac{\partial}{\partial \theta} -{\rm i} \theta \frac{\partial}{\partial t}
-{\rm i} \left(\psi_i-x_i \chi \right) \frac{\partial}{\partial x_i} \right),
\ee
hold invariant under the action of the $\mathcal{N}=1$ Schr\"odinger supergroup. Note that 
$\nabla_i$ is a Grassmann--even operator, $\nabla$ is a Grassmann--odd operator, and
$\frac{\partial}{\partial \theta} -{\rm i} \theta \frac{\partial}{\partial t}$ 
contributing to $\nabla$ is 
the conventional covariant derivative of the $d=1$, $\mathcal{N}=1$ supersymmetry 
subgroup. The latter anticommutes with the supercharge $Q$ displayed in (\ref{gener}) above.

The invariant Maurer--Cartan one--forms (\ref{MC}) along with the invariant derivatives (\ref{ID})
provide building blocks for constructing fluid mechanics with 
the $\mathcal{N}=1$ Schr\"odinger supersymmetry. Before formulating invariant 
equations, however, it proves convenient to implement a 
few field redefinitions and impose 
some constraints, which simplify the subsequent analysis. 

Firstly, 
analyzing the transformation rules
in (\ref{TRL}), one can unambiguously link $w$ to $u$, $\upsilon_i$, and $\chi$
\be\label{W}
w=\frac{1}{n-1} e^u \left(\frac{\partial \upsilon_i}{ \partial x_i} +\left(\frac{\partial}{\partial \theta} 
-{\rm i} \theta \frac{\partial}{\partial t} \right) \chi \right),
\ee
where, as above, $n$ designates the spatial dimension. Note that, as compared to the group--theoretic description 
of a perfect fluid with the Schr\"odinger symmetry in \cite{AG}, the 
Grassmann--odd superfield $\chi$ makes 
a contribution which modifies an overall factor in (\ref{W}). This latter fact is essential 
for implementing the bosonic limit below.

Secondly, imposing the
invariant constraints
\be\label{C1}
\omega_Q=0, \qquad \omega_{P i}=0,
\ee
and specifying to an orbit of a liquid superparticle 
$(x_i (t),\theta(t))$,
one can relate $\upsilon_i$ and $\chi$ to the velocity vector 
field and the velocity function associated with the Grassmann--even 
and Grassmann--odd coordinates, respectively 
\be
\frac{d x_i (t)}{ d t}=V_i (t,\theta (t),x(t)), \qquad
\frac{d \theta (t)}{ d t}=V (t,\theta (t),x(t)),
\ee
where
\be
V_i=\upsilon_i e^{{\rm i} \theta \chi}, \qquad 
V =-\chi.
\ee
The constraints (\ref{C1}) also allow one to represent the differential $d$ in the form
\be\label{MD}
d=d \tau \mathcal{D}, \qquad
\mathcal{D}=\frac{\partial}{\partial t}+ \upsilon_i (t,\theta,x) \frac{\partial}{\partial x_i}
-\chi (t,\theta,x) \left(\frac{\partial}{\partial \theta} 
-{\rm i} \theta \frac{\partial}{\partial t}  \right),
\ee
thus introducing an $\mathcal{N}=1$ 
supersymmetric generalization $\mathcal{D}$ of the conventional material 
derivative. 

Thirdly, the previous study of the perfect fluid equations with the Schr\"odinger symmetry in \cite{AG}
suggests a relation of
the scalar superfield $u$ to the fluid density $\rho$ via
\be
u=-\frac{2}{n} \ln \rho. 
\ee

Finally, the one--forms (\ref{MC}) suggest a simplifying field redefinition
\be
\psi_i \quad \to \quad \Sigma_i=\psi_i-x_i \chi. 
\ee
Having connected  $\upsilon_i$ and $\chi$ to the velocity 
superfields and $u$ to the density, one concludes that
$\Sigma_i$ is an extra Grassmann--odd vector superfield carried by a fluid, 
the presence of which is dictated by the structure of the superalgebra.

With these provisions, we are now in a position to formulate equations of fluid mechanics with 
the $\mathcal{N}=1$ Schr\"odinger supersymmetry. Imposing the constraint
\be
\omega_D=0, 
\ee
one obtains an analogue of the continuity equation
\be\label{EOM1}
\mathcal{D} u-\frac{2}{n-1} \left(\frac{\partial \upsilon_i}{\partial x_i}+\left(\frac{\partial}{\partial \theta} 
-{\rm i} \theta \frac{\partial}{\partial t}  \right) \chi \right)=0,
\ee
where $\mathcal{D}$ is the generalized material derivative in (\ref{MD}). It governs
evolution of the fluid density over time. 

Using the invariant one--forms $\omega_{K i}$, $\omega_H$ and 
the invariant Grassmann--even derivative
$\nabla_i$ in (\ref{ID}), one can build an analogue of the Euler equation, which determines the 
Grassmann-even velocity vector field $\upsilon_i$
\be\label{EOM2}
e^{\frac{u}{2}} \left(d \upsilon_i+{\rm i} \Sigma_i d \chi \right)=\alpha \nabla_i \omega_H 
\quad \rightarrow \quad 
\mathcal{D} \upsilon_i+{\rm i} \Sigma_i \mathcal{D} \chi+\alpha e^{-u} \frac{\partial u}{\partial x_i}=0,
\ee
where $\alpha$ is a Grassmann--even (coupling) constant. 

Similarly, from $\omega_S$, $\omega_H$, and 
the invariant Grassmann--odd derivative 
$\nabla$ in (\ref{ID}), one can build the equation of motion for the Grassmann--odd 
velocity function $\chi$
\be\label{EOM3}
e^{\frac{u}{2}} d\chi={\rm i} \beta \nabla \omega_H \quad \rightarrow \quad 
\mathcal{D} \chi+\beta e^{-u} \chi+{\rm i} \beta e^{-\frac{3u}{2}} \nabla u=0,
\ee
where $\beta$ is a Grassmann--even (coupling) constant. 

Finally, the invariant equation of motion for $\Sigma_i$ is built in terms of 
$\omega_{\Lambda i}$, $\omega_H$, $\nabla$, and $\nabla_i$ 
\bea\label{EOM4}
&&
d \Sigma_i+\chi \upsilon_i d\tau=-{\rm i} \gamma \nabla_i \nabla \omega_H-{\rm i} \sigma 
\nabla \nabla_i \omega_H \quad \rightarrow
\nonumber\\[2pt]
&&
\mathcal{D} \Sigma_i+\chi \upsilon_i+{\rm i} \gamma e^{-\frac{u}{2}} 
\left((\nabla u) \frac{\partial u}{\partial x_i}- \frac{\partial \nabla u}{\partial x_i} 
-\frac{{\rm i}}{2} e^{\frac{u}{2}} \chi \frac{\partial u}{\partial x_i} +{\rm i} e^{\frac{u}{2}} 
\frac{\partial \chi}{\partial x_i}
\right)
\nonumber\\[2pt]
&&
\qquad \qquad \qquad
+{\rm i} \sigma e^{-\frac{u}{2}} 
\left(\frac{1}{2} (\nabla u) \frac{\partial u}{\partial x_i}-\nabla \left(  
\frac{\partial u}{\partial x_i} \right)
-{\rm i} e^{\frac{u}{2}} \chi \frac{\partial u}{\partial x_i}\right)=0,
\eea
where $\gamma$ and $\sigma$ are Grassmann--even (coupling) constants. To summarize,
(\ref{EOM1})--(\ref{EOM4}) determine fluid
mechanics with the $\mathcal{N}=1$ Schr\"odinger supersymmetry within the group--theoretic approach.

A few comments are in order. Firstly, the consideration above demonstrates that the 
group--theoretic construction in \cite{AG} can be generalized to accommodate $\mathcal{N}=1$ 
supersymmetry. On the technical side, 
essential new ingredients were the invariant Grassmann--odd derivative 
$\nabla$ in (\ref{ID}) and the fermionic contribution to the superfield
$w$ in (\ref{W}).

Secondly, considering the component decompositions\footnote{We use the conventional notation for 
complex conjugation of two Grassmann--odd
variables $\theta_1$ and $\theta_2$: 
${\left(\theta_1 \theta_2 \right)}^{*}=\theta_2^{*} \theta_1^{*}$. Recall that all fermions 
considered in this work are chosen to be real, 
e.g. ${\left(\theta \right)}^{*}=\theta$.}
\bea\label{comp}
&&
u(t,\theta,x)=u_0 (t,x)+{\rm i} \theta u_1 (t,x), \qquad \upsilon_i (t,\theta,x)=\upsilon_{0 i} (t,x)+
{\rm i} \theta \upsilon_{1 i} (t,x),
\nonumber\\[2pt]
&&
\chi (t,\theta,x)=\chi_1 (t,x)+\theta \chi_0 (t,x), \qquad 
\Sigma_i (t,\theta,x)=\Sigma_{1 i} (t,x)+
\theta \Sigma_{0 i} (t,x),
\eea
where Grassmann--even components are labeled by the subscript $0$ and Grassmann--odd 
components are marked by the subscript $1$, one concludes that each component field in 
question has a superpartner of 
the opposite Grassmann parity. Similarly, each equation of motion in (\ref{EOM1})--(\ref{EOM4}) yields a pair of 
component equations. For example, (\ref{EOM1}) splits into
\bea
&&
\left(\frac{\partial}{\partial t} +\upsilon_{0 i} \frac{\partial}{\partial x_i} \right) u_0
-\frac{2}{n-1} \frac{\partial \upsilon_{0 i}}{\partial x_i}-{\rm i} \chi_1 u_1-\frac{2}{n-1} \chi_0=0,
\nonumber\\[2pt]
&&
\left(\frac{\partial}{\partial t} +\upsilon_{0 i} \frac{\partial}{\partial x_i} \right) u_1+
\upsilon_{1 i} \frac{\partial u_0 }{\partial x_i}
-\frac{2}{n-1} \frac{\partial \upsilon_{1 i}}{\partial x_i}-\chi_0 u_1 
-\chi_1 \frac{\partial u_0}{\partial t}+\frac{2}{n-1} \frac{\partial \chi_1}{\partial t}=0.
\eea
While the Grassmann--even components $u_0 (t,x)$ and $\upsilon_{0 i} (t,x)$ 
bear clear physical meaning (they link to the density and the velocity vector field of an actual fluid), 
interpretation of other components is somewhat dubious. Note that this problem is typical 
for all the superfield formulations 
of fluid mechanics (see e.g. \cite{DP,FKT,GH}). Yet, although the Grassmann--odd components in (\ref{comp}) 
can hardly be given a physical interpretation, their
quadratic combinations are Grassmann--even and might be of potential interest. 

Thirdly, the bosonic limit of Eqs. (\ref{EOM1})--(\ref{EOM4}) is more subtle as it may seem at first glance. 
Comparing the group--theoretic element (\ref{gte}) with a similar object relevant for describing a perfect fluid with 
the Schr\"odinger symmetry \cite{AG} 
\be\label{gteq}
g=e^{{\rm i} t H}  
e^{{\rm i} x_i P_i} 
e^{{\rm i}  \upsilon_{0 i} (t,x) K_i} 
e^{{\rm i} u_0 (t,x) D} e^{{\rm i} w_0 (t,x) C},
\ee
and taking into account the component decompositions (\ref{comp}), one concludes that 
\be 
\theta  \to 0, \qquad \chi_1 (t,x)  \to  0, \qquad \psi_{1 i} (t,x) \to 0,
\ee
suffice to determine the bosonic limit. Note that other components in (\ref{comp}) may remain arbitrary.
In particular, the perfect fluid equations with the Schr\"odinger symmetry follow from (\ref{EOM1})--(\ref{EOM4}) 
provided $\chi_0 \to -\frac{1}{n} \frac{\partial \upsilon_{0 i}}{\partial x_i}$, which 
is consistent with the transformations rules of the components under the Schr\"odinger subgroup.

Fourthly, one of the motivations for the present work was to verify whether the bosonic 
component 
of the extra vector superfield $\Sigma_i$, the presence of which is dictated by the superalgebra,
might contribute to the bosonic limit of the equation of motion for $\upsilon_i$. 
Potentially, this might provide an extension of the conventional Euler 
equation by extra contributions, which would describe couplings to an external vector field. 
Unfortunately, this did not prove to be the case.

\vspace{0.5cm}

\noindent
{\bf 5. Conclusion}\\

To summarize, in this work the group--theoretic description of nonrelativistic
conformally invariant fluids in \cite{AG} has been extended to cover the case of the 
$\mathcal{N}=1$ Schr\"odinger supersymmetry. Specifically, a group--theoretic element was
introduced in which coordinates parameterizing a superspace were associated with 
the generators of (super)translations, while other generators in the superalgebra
were accompanied by arbitrary superfields of the same Grassmann parity. 
Transformation rules of the 
coordinates and fields under the $\mathcal{N}=1$ Schr\"odinger supergroup 
were established and then used to identify
some of the fields with those characterizing a fluid.
The invariant Maurer--Cartan one--forms 
and two invariant derivatives were constructed, which gave rise to equations of motion of a fluid mechanics
with the $\mathcal{N}=1$ Schr\"odinger supersymmetry.

Turning to possible further developments, it would be interesting to apply the 
solution generating techniques in \cite{GH} to the set 
of equations (\ref{EOM1})--(\ref{EOM4}). A Hamiltonian formulation reproducing  (\ref{EOM1})--(\ref{EOM4})
is worth studying as well. Finally, an extension of the 
present analysis to the case of the $\ell$--conformal Galilei supersymmetry and
the Lifshitz supersymmetry might be of some potential interest. 

\vspace{0.5cm}

\noindent{\bf Acknowledgements}\\

\noindent
This work was supported by the Russian Science Foundation, grant No 23-11-00002.


\begin{thebibliography}{nn}
\bibitem{JP}
R. Jackiw, A.P. Polychronakos, {\it Supersymmetric fluid mechanics}, 
Phys. Rev. D {\bf 62} (2000) 085019, hep-th/0004083.
\bibitem{BJ}
Y. Bergner, R. Jackiw, {\it Integrable supersymmetric fluid mechanics from superstrings},
Phys. Lett. A {\bf 284} (2001) 146, physics/0103092.
\bibitem{DP}
A. Das, Z. Popowicz, {\it Supersymmetric polytropic gas dynamics}, Phys. Lett. A {\bf 296} (2002) 15, hep-th/0109223.
\bibitem{FKT}
B.W. Fatyga, V.A. Kostelecky, D.R. Truax,  {\it Grassmann--valued fluid dynamics}, J. Math. Phys. {\bf 30}
(1989) 1464.
\bibitem{GH}
A.M. Grundland, A.J. Hariton, {\it Supersymmetric version of the Euler system and its 
invariant solutions}, 
Symmetry {\bf 5} (2013) 253.
\bibitem{BJLNP}
B. Bistrovic, R. Jackiw, H. Li, V.P. Nair, S.-Y. Pi, {\it Non-abelian fluid dynamics in Lagrangian formulation}, Phys. Rev. D {\bf 67} (2003) 025013, hep-th/0210143.
\bibitem{JNPP}
R. Jackiw, V.P. Nair, S.Y. Pi, A.P. Polychronakos, {\it Perfect fluid theory and its extensions}, 
J. Phys. A {\bf 37} (2004) R327, hep-ph/0407101.
\bibitem{CWZ}
S.R. Coleman, J. Wess, B. Zumino, {\it Structure of phenomenological Lagrangians. I}, Phys. Rev. {\bf 177} (1969) 2239.       
\bibitem{AG}
A. Galajinsky, {\it Group--theoretic approach to perfect fluid equations with 
conformal symmetry}, Phys. Rev. D {\bf 107} (2023) 026008, arXiv:2210.14544.
\bibitem{HH}
M. Hassaine, P.A. Horvathy, {\it Field--dependent symmetries of a non--relativistic fluid
model}, Annals Phys. {\bf 282} (2000) 218, math-ph/9904022.
\bibitem{RS}
L. O'Raifeartaigh, V.V. Sreedhar, {\it The maximal kinematical invariance group of fluid
dynamics and explosion-implosion duality}, Annals Phys. {\bf 293} (2001) 215, hep-th/0007199.
\bibitem{HH1}
M. Hassaine, P.A. Horvathy, {\it Symmetries of fluid dynamics with polytropic exponent},
Phys. Lett. A {\bf 279} (2001) 215, hep-th/0009092.
\bibitem{HZ}
P.A. Horvathy, P.-M. Zhang, {\it Non--relativistic conformal symmetries 
in fluid mechanics}, Eur. Phys. J. C {\bf 65} (2010) 607, arXiv:0906.3594.
\bibitem{GGT}
J.P. Gauntlett, J. Gomis, P.K. Townsend, {\it Supersymmetry and the physical phase space
formulation of spinning particles}, Phys. Lett. B {\bf 248} (1990) 288.
\bibitem{BG}
K. Balasubramanian, J. McGreevy, {\it Gravity duals for
non-relativistic CFTs}, Phys. Rev. Lett. {\bf 101} (2008) 061601, arXiv:0804.4053.
\bibitem{MR}
M. Rangamani, {\it Gravity and hydrodynamics: Lectures on the fluid-gravity correspondence}, 
Class. Quant. Grav. {\bf 26} (2009) 224003, arXiv:0905.4352.
\bibitem{DH}
C. Duval, P.A. Horvathy, {\it On Schrodinger superalgebras}, J. Math. Phys. {\bf 35} (1994) 2516,
hep-th/0508079.
\end{thebibliography}
\end{document}